# Effects of annotation granularity in deep learning models for histopathological images

Jiangbo Shi[1,2,3], Zeyu Gao[1,2,3], Haichuan Zhang[1,2,3], Pargorn Puttapirat[1,2,3], Chunbao Wang[4], Xiangrong Zhang[5], Chen Li[1,2,3*]
[1] National Engineering Lab for Big Data Analytics Xi'an Jiao tong University, Xi'an, Shaanxi 710049, China
[2] School of Electronic and Information Engineering, Xi'an Jiao tong University, Xi'an, Shaanxi 710049, China
[3] Shaan xi Province Key Laboratory of Satellite and Terrestrial Network Tech.R&D,
Xi'an Jiao tong University, Xi'an, Shaanxi 710049, China
[4] Department of Pathology, the First Affiliated Hospital of Xi'an Jiaotong University, Xi'an, Shaanxi 710061, China
[5] Institute of Intelligent Information Processing, Xidian University, Xi'an, Shaanxi 710071, China
Email: {shijiangbo, gzy4119105156, zhanghaichuan, pargorn}@stu.xjtu.edu.cn, bingliziliao2012@163.com,
xrzhang@mail.xidian.edu.cn, cli@xjtu.edu.cn

*Abstract*—Pathological examination is an important step in cancer diagnosis. Pathologists make diagnosis and pathology report based on observed cell and tissue structure on pathological slides. With the development of statistical machine learning, especially deep learning, automated classifiers are being used to analyze histopathological slides and assist pathologists in diagnosis. Currently, commonly used annotation methods in histopathological slides include image-wise, bounding box, ellipse-wise, pixel-wise. In order to verify the influence of annotation in pathological slide on deep learning model, we design corresponding experiments to test the performance based on annotations with different granularity annotation. In classification, all state-of-the-art deep learning based classifiers perform better when they are trained by the dataset with pixel-wise annotation. On average, precision, recall and F1-score improves by 7.87%, 8.83% and 7.85% respectively. Thus, it is suggested that finer annotations are better utilized by deep learning algorithms in classification tasks. Similarly, semantic segmentation algorithms can achieve 8.33% better segmentation accuracy when trained by pixel-wise annotations. Our study shows that finer-grained annotation can not only improve the performance of deep learning models, but help deep learning model extract more accurate phenotypic information from histopathological slides. An accurate acquisitions of phenotypic information can help pathologists to enquire the model based on which regions and features in the slide were mainly used to calculate the prediction, improve the reliability of the model prediction. The compartmentalized prediction approach similar to this work may contribute to phenotype and genotype association studies.

## I. INTRODUCTION

Nowadays, cancer is the biggest public health problem in the world. The global caner burden is estimated to have risen to 18.1 million new cases and 9.6 million deaths in 2018, and there will be 2.2 million new cases in 2020 [29]. About 80%-90% early cancer patients can be cured. Therefore, early diagnosis and treatment can improve the survival rate and the patient quality of life. Pathological examination is the gold standard in cancer screening and has the highest reliability. During the slide reviewing process, clinicians usually obtains small pieces of diseased tissue from the patient through a biopsy, then pathological slides are prepared. Pathologists will perform a microscopic examination under a microscope to give a detailed pathological diagnosis report. Pathological report generally contains these following information, such as the extent of tumor spreading, the grade and source of cancer cells and the degree of malignancy. According to the pathological report given by pathologist, clinicians will develop a corresponding treatment plan in order to provide the best of care to improve the cure rate and prognosis of the patient. As the number of cancer patients around the world continues to increase, the workload of pathologists is getting heavier. From 2007 to 2017, the number of pathologists per 10,000 people in United States dropped from 5.16 to 3.94, and the diagnostic workload of pathologists increased by 41.73% [30]. Pathology slide reviewing is a time-consuming and laborious work that relies on a manual qualitative analysis by one or more pathologists. Moreover, the diagnosis between different pathologists are prone to disagreement. This is not only related to the subjective judgment of the pathologists, but also affected by the environment. The emergence of medical imaging equipment has enabled traditional glass slides to be imaged by a scanner, thus a high quality whole slide images(WSIs) can be obtained. The accumulation of WSIs make it possible to implement computer-assisted diagnosis. Moreover, with the development of deep learning in the fields of computer vision and image processing in recent years, deep learning can be used to achieve tasks such as classification, detection and segmentation on digital slides to help pathologists obtain some of pathological indicators, assist pathologists to conduct pathological diagnosis, improve the efficiency and accuracy of pathological examination.

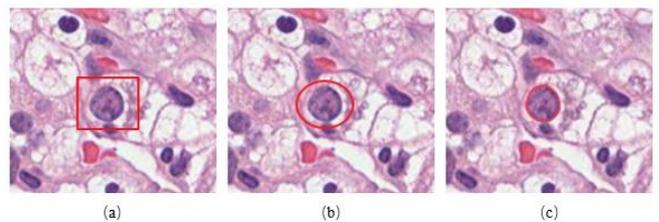

Fig. 1. Different annotation granularity. including (a) bounding box, (b) ellipse-wise, (c) pixel-wise granularity annotation.

The training of deep learning models relies on large-scale annotated data. Currently, a few pathological image datasets have been published [18-20]. These datasets have different granularity of annotation, including image-wise, bounding box, ellipse-wise, pixel-wise as demonstrated in Fig. 1. For example, all singnet ring cell are annotated using bounding box in the DigestPath 2019 competition [20]. Different granularity of dataset has the different characteristics for showing the pathological images. First, finer-grained granularity annotation can annotate more precise features on pathological image, such as morphological features, color and texture features. These precisely annotated features can better help train deep learning models and improve the performance of deep learning models. In addition, although there are some work currently using deep learning to assist pathologists in pathological diagnosis, due to the unexplained characteristic of deep learning, if the relevant areas are not accurately annotated, the models can't tell the pathologists that it is based on what features or regions to give the final prediction results. The interpretability of the model plays an import role in ensuring the performance of the model and the reliability of the application in actual clinical diagnosis. Moreover, fine granularity annotation can accurately display cells phenotype on histopathological image. Accurate access to phenotypic information of cells has a great help for genotype-phenotype association study.

In order to verify the impact of the granularity annotation on the performance of deep learning models. We design two experiments. The first one is that we design and conduct a patch-based CNN classification experiment. We build corresponding dataset for each granularity annotation. Then, we perform same experiment on each dataset with different granularity annotation. The classification model trained by pixel-wise granularity annotation outperforms other two kinds of granularity annotation. The second one is semantic segmentation which can acquire the morphological features of cancer cells on slides. We use full convolutional neural network to do the semantic segmentation experiment. Mask generated by the model which is trained on finer-grained granularity annotation can accurately display the contour information and determine the grade of cancer cells.

## II. RELATED WORK

Histopathological images carry informative cellular phenotypes, however, they also contain a large amount of complex and redundant information. Therefore, an important step in pathological analysis is to extract meaningful visual features from slides. Traditional machine learning methods rely on manually engineered features. The manually engineered feature refers to selecting and simplifying the low-dimensional vectors that best express the image content, including gray histograms, shape features, texture features, and relationship with surrounding tissues. However, these features have the following defects, First, the selection are mainly dependent on the experts, with poor objectivity. They may not be able to characterize and extract the comprehensive information presented in the slides. Second is lack of principle standard to merge different kinds of mutual features. Deep learning can learn the expression of the invariance and deformation insensitivity from a number of training data, which can fully express different features on slide without being restricted by professional factors. At present, the analysis of pathological images using deep learning mainly includes two aspects: classification at organizational level or grading of cancer, segmentation of cells or tissues.

Pathological image reflects the biological behavior and morphological characteristics of the tissue cells. Pathological grade reflects the morphological difference between tumor tissue and normal tissue cell in tissue structure and cell morphology, and can be used to determine whether the tissue is cancerous. Due to its local perception and parameter sharing characteristics, convolutional neural networks can automatically extract features from slides and implicitly learn from training data. In recent years, new progress has been made in the use of CNN for image classification. Roy et al [1] trained a patch-based convolutional neural network model to automatically classify breast cancer pathological image into four categories, classified as normal, benign, in situ, and invasive carcinoma. The accuracy is 84.7% as patch-wise level, and at image-wise level, an accuracy of 92.5% was obtained. Xie et al [2] trained two classification models RetNet50 [3] and VGG19 [4] for melanoma cancer, and achieved good results in a two-class task. Xia et al [5] pre-trained a GoogleNet model in Lymph node metastases, and use this model to make predictions on prostate cancer dataset with less data. They got an accuracy of 84.3%. Arvaniti et al [6] adopted Gleason grading standard, and trained a MobileNet model on prostate cancer dataset. The macro-average recall is 70% in a four classification problem. Therefore, relevant researchers have obtained good experiments results in the corresponding fields by using CNN-based methods in the classification of pathological images, which reflects the universality of CNN models in pathological image classification.

Segmentation of nuclei has been addressed by many authors with a variety of traditional approaches, most of which are based on active contours [7], region growing [8], intensity thresholding [9] and watershed [10]. With Convolutional Neural Networks (CNN) showing its ability to classify a single pixel, based on its neighborhood described by high-level features, CNN is widely used in nuclei detection n and segmentation [11-13]. Yan Xu [14] et. al segment individual glands in colon histology images with the image instance segmentation method based on fully convolutional networks (FCN). Ling Zhang [15] et. al propose a novel approach for segmentation of cervical nuclei that combines FCN and graph-based approach (FCNG). Peter Naylor [16] et. al address the problem of segmenting touching nuclei by formulating the segmentation problem as a regression task of the distance map and demonstrate superior performance of this approach as compared to other approaches using CNN. Amirreza Mahbod [17] et.al propose a novel approach to segment touching nuclei in H&E-stained microscopic images using U-Net based models in two sequential stages.

Whether it is a classification task or a segmentation task, a dataset of related pathological images is needed to train the model. Currently, there have been some pathological image datasets published. Two datasets named BACH [18] related to breast cancer is published in the ICIAR 2018 Grand Challenge On Breast Histology Images. The first dataset is microscopic images, the size of each image is 2048 by 1536, using image-wise annotation, only label given to each microscopic image. The image data in the second dataset are WSIs with pixel-wise annotation. The pathologists use manually methods to mark the cancerous tissue area on the slide. CAMELYON 16 competition [19] publishes a sentinel lymph node dataset with 400 WSI. In this dataset, pathologist uses pixel-wise annotation to separate the normal tissue and infiltration accurately. There are another two pathological datasets which are published in DigestPath 2019 [20]. First dataset is about Signet ring cell which includes 700 pathological slides collected from 120 patients. The slide is stained with H&E and scanned at 40X magnification. Professional pathologist uses bounding box annotation to annotate all the signet ring cell. The second is colonoscopy tissue segment dataset which collects 1000 pathological from 700 patients. All colonoscopy tissues are labeled by a professional pathologist using pixel-wise annotation.

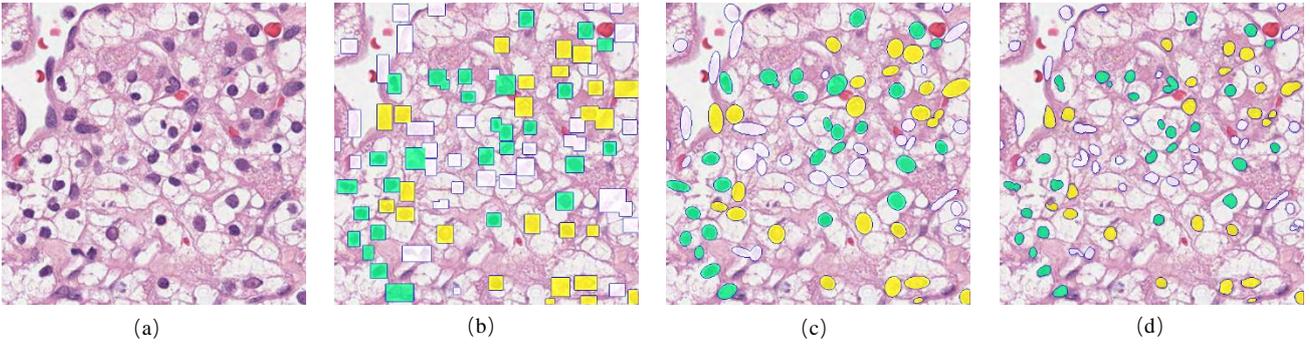

Fig. 2. Different granularity annotation using in our dataset, (b),(c),(d) represent bounding box, ellipse-wise, pixel-wise granularity annotation. White color annotation represents the normal cell nucleus, green and yellow color annotation represent grade 1 and grade 2 cancer cell nucleus.

TABLE I. WHO/ISUP GRADING STANDARD

| Grade | Description |
|---|---|
| Grade 1 | Tumor cell nucleoli invisible or small and basophilic at 400 x magnification |
| Grade 2 | Tumor cell nucleoli conspicuous at 400 x magnification but inconspicuous at 100 x magnification |
| Grade 3 | Tumor cell nucleoli eosinophilic and clearly visible at 100 x magnification |
| Grade 4 | Tumors showing extreme nuclear pleomorphism and/or containing tumor giant cells and/or the presence of any proportion of tumor showing sarcomatoid and/or rhabdoid dedifferentiation |

## III. MATERIALS AND METHODS

### A. Dataset

The visual features of ccRCC are clear with less distracting information which is advantageous to high granularity annotation. So we use clear cell renal carcinoma cancer(ccRCC) as research object to investigate the impact of different granularity annotation on the performance of deep learning models in classification and semantic segmentation tasks. TCGA [9] is a project jointly supervised by the National Cancer Institute and the American Human Genome Research Institute. This database contains 60 tissues, 38 cancers of organs and their subtypes. Our team downloads 10 whole slide images of ccRCC from TCGA database. ccRCC is divided into 4 grades based on the ISUP grading standard as shown in table 1. We just consider two kinds of cancer cells in our work which is grade1 and grade 2. we choose twenty diagnostic areas at 20X magnification on each slide. The size of each selected region is 800*800 pixels. Six rigorously trained annotators use OpenHI [21] to annotate the chosen regions according to ISUP grading standard [22]. Six annotators are divided into two groups. First group annotates all the odd regions. second group annotates all the remaining regions. After completing the data annotation, we compute the inter-rater reliability between different annotators in a group. In the first group, the average kappa statics value [23] is 0.7137. another group's kappa statics value is 0.7092. According to the interpretability of kappa statics, the confidence of different annotators is at moderate level. This confidence is acceptable because of the complex phenotypic information on the pathological slides. In order to improve the quality of annotation, we combine three annotator's annotations in one group by selecting the annotation that most people agree on. Finally, we get 200 slides with accurate pixel-wise annotations. Later, we utilize OpenCV to expand pixel-wise annotation into ellipse-wise and bounding box annotation. We build the corresponding dataset for each granularity of annotation as shown in Fig. 2.

### B. Classification framework

For the classification task, the framework mainly includes three parts: data processing, model selection and training, evaluation metrics. We use the same processing method for each granularity dataset to compare the impact of different granularity annotation on the performance of deep learning model in classification task.

#### 1) Data preprocessing

##### a) Patch generation

Pathological images generally have large size, so we need to cut the images into patches. In order to use CNN to do the classification task and verify the consistency of the granularity annotation impact under different window size, we use three sizes of sliding windows. The patch size is 32*32, 64*64, 128*128 separately. The stride is the half of corresponding sliding window.

##### b) Patch label determination

There are maybe different kinds of annotation in a patch. Therefore, we need to determine the label of each patch according to certain criteria. If a patch just contains one kind of annotation, the label corresponding to the annotation is used as the label of the patch. If a patch contains more than one kind of annotation, we choose the highest grade label as the patch label. If there are not a complete nucleus in a patch, we just discard this kind of patch.

##### c) Data augmentation and selection

Unbalanced and limited data size is the major challenge in the development of robust computer-aided diagnosis (CAD) system. Data augmentation is an approach used in deep learning models to enlarge the dataset in order to alleviate the problem of limited data size. the selection of data augmentation approach should be performed wisely, based on the dataset. To augment the dataset, we rotate each patch by 4 multiples of 90°, with and without mirroring, which results in 8 valid variations for each patch. We further apply random color perturbations to these variations as suggested by [24] and produce 8 more patches. The color augmentation process would help our model to learn color-invariant features and make pre-processing color normalization [25]. In order to ensure the balance of different kinds of patches, the same number of patches are selected for each grade of patch to build the final dataset.

*2) Mode selection and training*

We choose three deep convolutional neural networks ResNet18 [3], VGG16 [4], MobileNet [26] to do the classification task.

ResNet is the first place in the ILSVRC 2015 competition. The main idea is to add skip connection structure to the network. By directly putting the input information to the output to protect the integrity of the information, the entire network only needs to learn the different between input and output, simplifying the learning objectives and difficulty. The network is thinner and controls the number of parameters; using fewer pooling layers and a large number of downsampling to improve the propagation efficiency; using batch normalization and global average pooling for regularization to speed up the training; reducing the number of 3*3 convolutions and using more 1*1 convolutions when the number of layers is high. We choose ResNet18 as one of the classification models.

VGG was proposed by Oxford's Visual Geometry Group, which won second place in the 2014 ILSVRC competition. VGG has two structures: VGG16 and VGG19. We choose VGG16 for experiments in this paper. VGG uses several 3*3 convolution kernels in their model instead of the larger convolution kernels in AlexNet(11*11, 7*7, 5*5). For a given receptive field, the multi-layer nonlinear can increase the network depth to ensure the learning of more complex patterns and the cost is relatively small. VGG has a very deep network hierarchy, which includes 16 hidden layers(13 convolutional layers and 3 fully connected layers). The entire network uses the same convolution kernel size 3*3 and maximum pool size 2*2.

MobileNet was originally proposed by Google in 2017. It is a small and efficient CNN model for the scenarios that require low latency. The basic unit of MobileNet is depthwise separable convolution which is a factorized convolution. It can be divided into two smaller operations: Depthwise convolution and pointwise convolution. Depthwise convolution uses different size of convolution kernel according to different input channels. Pointwise convolution uses 1*1 convolution kernel. For depthwise separable convolution, first is to use depthwise convolution to convolve separately for different input channels, and then use the pointwise convolution to combine the above output. This over effect is similar to a standard convolution, but greatly reduces the amount of calculations and model parameters. The network structure is: first is a 3*3 standard convolution, then stacking depthwise separable convolution, then through the average pooling layer, and finally a softmax layer, the entire network has s total of 28 layers.

These three classification models are pretrained on ImageNet dataset. We fine-tuned selected model from ImageNet-learned parameters. We use the Adam optimization technique with initial learning rate of 0.001 and Nesterov momentum 0.9. For all three models, the categorical cross-entropy loss is used as minimization objective function. Image patches are resized to 128*128 for each size of patch. We use 80% of the samples for training, 10% of the samples for testing and 10 % the samples for evaluation.

*3) Evaluation metrics*

For patch-based CNN tasks, we use accuracy, recall, precision, and F1 score as evaluation metrics. The formula of accuracy, precision and recall are shown in (1), (2) and (3).

(TP: true-positive; TN: true-negative; FP: false-positive; FN: false-negative)

$$\text{Accuracy} = \frac{TP + TN}{TP + TN + FP + FN} \quad (1)$$

$$\text{Precision} = \frac{TP}{TP + FP} \quad (2)$$

$$\text{Recall} = \frac{TP}{TP + FN} \quad (3)$$

*C. Semantic segmentation framework*

For semantic segmentation, the framework also includes three parts: data preprocessing, model selection and training, evaluation metrics.

*1) Data preprocessing*
   *a) Image with mask*

For semantic segmentation task, the data fed into the selected model is the original pathological slide and corresponding mask with different granularity annotation.

   *b)* Data augmentation

We use the same data augmentation methods in semantic segmentation task as the classification experiment. The same augmentation operation was used for pathological images and corresponding masks.

*2) Model selection and training*

In the semantic segmentation task, we choose three fully convolutional neural network models: UNet [13] FPN [28], LinkNet [29].

In medical image segmentation tasks, the most commonly used model is UNet. UNet uses a structure named encoder-decoder. The entire model is divided into two parts. The first half for feature extraction and the second half for upsampling. Compared to other common segmentation networks, UNet uses a completely different feature fusion method: concatenate, avoiding the supervision and loss calculation directly in the advanced feature map, but using many low-level features. Therefore, the resulting feature map not only includes the high-level features, but also contains many low-level features, which achieves the fusion of features at different scales and improves the accuracy of the model results. For pathological image, the pathologist needs both information of the whole picture to judge the global information such as position, the contrast of the lesion and the normal position, and also the local information of some positions, so it is necessary to consider at multiple scales as much as possible. UNet can combine high and low level feature to increase information content.

FPN(feature pyramid network) is a method to efficiently extract features of each dimension in a slide by using CNN model. Feature extraction is divided into three parts: generation of different dimensional features from top to bottom, feature enhancement from top to bottom, the correlation expression between the features of CNN network layer and each dimension of the final output. This feature of FPN greatly improves the detection performance of small objects.

LinkNet directly connects the encoder to the decoder to increase accuracy. The LinkNet architecture is similar to a ladder network structure in which the feature map of the encoder and the upsampled feature map of the decoder are added. Due to the channel reduction scheme, the decoder module contains quite a few parameters.

TABLE II. CLASSIFICATION RESULT

| patch size | evaluation metric | Vgg16 | | | ResNet18 | | | MobileNet | | |
|---|---|---|---|---|---|---|---|---|---|---|
| | | Bb | Ellipse | Pixel | Bb | Ellipse | Pixel | Bb | Ellipse | pixel |
| 32 | Precision | 85.49% | 88.59% | **92.04%** | 87.68% | 89.88% | **93.55%** | 87.27% | 89.47% | **94.25%** |
| | Recall | 83.47% | 88.79% | **92.30%** | 88.05% | 90.11% | **93.73%** | 87.65% | 89.83% | **94.40%** |
| | F1-score | 0.8485 | 0.8864 | **0.9214** | 0.8779 | 0.8997 | **0.9362** | 0.8741 | 0.8962 | **0.9432** |
| 64 | Precision | 92.08% | 93.38% | **94.77%** | 88.03% | 89.21% | **92.03%** | 83.85% | 86.96% | **91.06%** |
| | Recall | 92.09% | 93.40% | **94.77%** | 88.07% | 89.24% | **92.08%** | 83.93% | 87.03% | **91.13%** |
| | F1-score | 0.9207 | 0.9338 | **0.9477** | 0.8804 | 0.8920 | **0.9204** | 0.8380 | 0.8695 | **0.9106** |
| 128 | Precision | 79.79% | 85.09% | **87.66%** | 74.95% | 78.22% | **79.55%** | 77.31% | 78.84% | **80.16%** |
| | Recall | 79.83% | 85.04% | **87.65%** | 74.99% | 78.06% | **79.63%** | 77.36% | 78.88% | **80.20%** |
| | F1-score | 0.7980 | 0.8506 | **0.8765** | 0.7495 | 0.7806 | **0.7956** | 0.7728 | 0.7885 | **0.8016** |

TABLE III. CLASSIFICATION RESULT FOR EACH GRADE OF CELL NUCLEUS

| patch size | evaluation metric (accuracy) | Vgg16 | | | ResNet18 | | | MobileNet | | |
|---|---|---|---|---|---|---|---|---|---|---|
| | | Bb | Ellipse | Pixel | Bb | Ellipse | Pixel | Bb | Ellipse | pixel |
| 32 | Grade0 | 89.03% | 94.07% | **95.99%** | 92.38% | 94.52% | **96.68%** | 92.30% | 93.96% | **96.90%** |
| | Grade1 | 65.51% | **67.53%** | 67.29% | 74.48% | 72.93% | **75.15%** | 74.43% | 74.04% | **78.92%** |
| | Grade2 | 61.30% | **65.61%** | 65.43% | **72.27%** | 68.59% | 72.08% | 68.00% | 66.79% | **73.86%** |
| 64 | Grade0 | 92.92% | 94.54% | **96.55%** | 90.20% | 91.37% | **93.82%** | 85.33% | 89.55% | **93.27%** |
| | Grade1 | 91.47% | **93.23%** | 93.08% | 86.94% | 87.20% | **90.04%** | 82.97% | 84.43% | **88.34%** |
| | Grade2 | 91.18% | 90.28% | **91.28%** | 84.82% | 86.72% | **89.09%** | 81.92% | 84.23% | **87.94%** |
| 128 | Grade0 | 72.33% | 78.28% | **84.41%** | 67.65% | 70.80% | **74.61%** | 70.25% | 72.02% | **76.93%** |
| | Grade1 | 81.22% | 87.09% | **89.12%** | 77.81% | 81.05% | **81.20%** | 77.62% | 80.63% | **81.71%** |
| | Grade2 | 80.87% | 85.58% | **87.48%** | 74.23% | 78.52% | **79.61%** | 79.58% | 79.56% | **79.84%** |

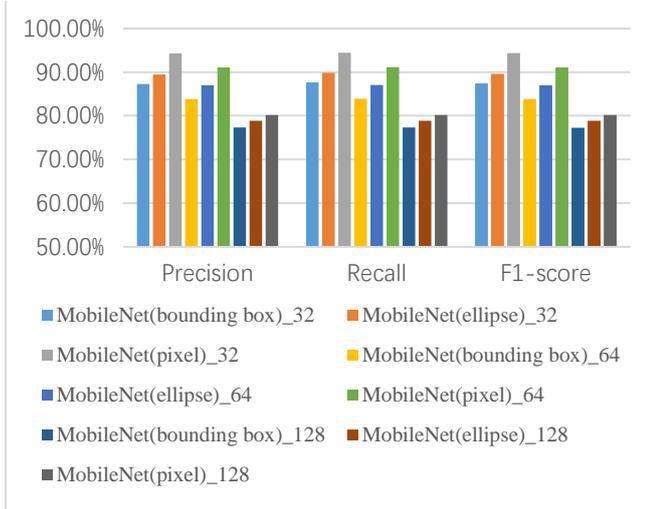

Fig. 3. Classification result predicetd by MobileNet under three different granularity annotation and three patch sizes. Bounding box, ellipse, pixel represent three different granularity anoattion. 32, 64, 128 represent different sizes of patch.

*3) Evaluation metrics*

We choose accuracy, dice score, nuclei recall, cancer recall as evaluation metrics. Accuracy represents the accuracy of each pixel classification. The computational formula of dice score is as (4). $V_{seg}$ represents predicted result of the model. $V_{gt}$ represents the ground truth.

$$\text{DICE} = \frac{2*(V_{seg}\ and\ V_{gt})}{V_{seg} + V_{gt}} \quad (4)$$

Nuclei recall represents the recall rate of nuclei in the predicted result. Specifically, the computation is that the sum of the number of pixels that are correctly predicted as the nucleus and number of pixels that are incorrectly predicted as the nucleus is divided by number of pixels that are correctly predicted as the nucleus.

Cancer recall represents the recall rate of cancer cell. Specifically, the computation is that the sum of the number of pixels that are correctly predicted as the cancer cell nucleus and the number of pixels that are incorrectly predicted as the cancer cell nucleus is divided by the number of pixels that are correctly predicted as the cancer cell nucleus.

RESULT

*A. Classification task*

The results of classification experiment are shown in Table 2 and Table 3. In Table 2, three models Vgg16, ResNet18 and MobileNet are trained under three kinds of granularity annotation and three different size of patches. The pixel-wise granularity annotation obtains the best experiment results in each case. Precision, recall and F1-score improves by 7.87%, 8.83% and 7.85% respectively. In Fig.3, we can clearly see that pixel-wise annotation has an improvement than other two kinds of granularity annotation in three patch size. The main reason is that finer-grained granularity annotation can build more accurate patches in the preprocessing step, improve the quality of dataset, and then affect the performance of model. In Table 3, grade 0 represents normal cell nucleus, grade 1 and grade 2 represent two different grades of cancer cell nucleus. For each types of nuclei, three models trained by pixel-wise annotation almost obtain the best accuracy comparing with bounding box and ellipse-wise granularity annotation in each case. Moreover, for grade 1 and grade 2, patch size with 64 gets the highest accuracy than other size of patches. The reason is that the patch size with 32 only focuses on the features of nucleus. The patch size with 128 contains more mesenchyme in one patch, which makes the features contained in one patch more complicated. The model cannot extract accurate features from the patch. In a word, finer-grained granularity annotation can help improve the performance of deep learning model.

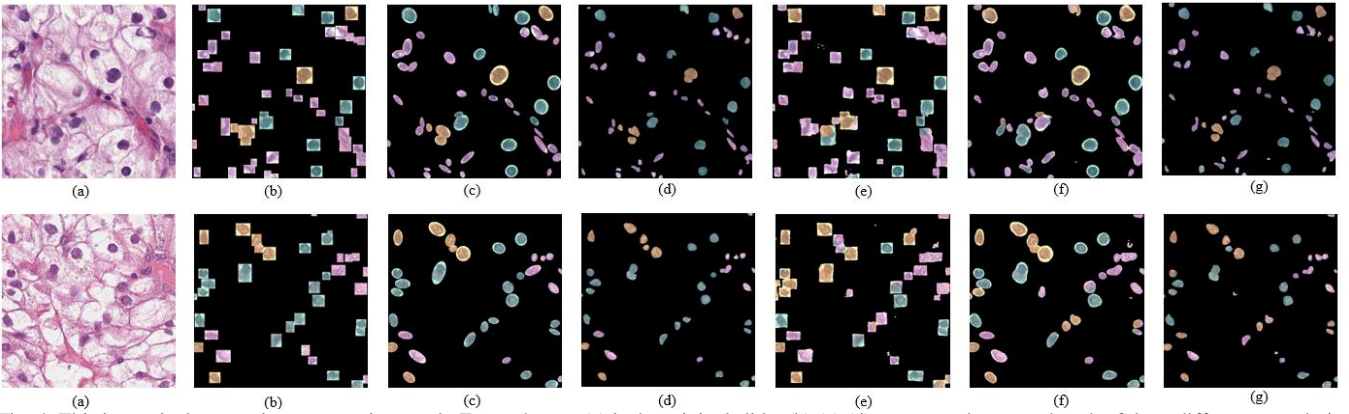

Fig. 4. This image is the sematic segmentation result. For each row, (a) is the original slide, (b),(c),(d) represent the ground truth of three different granularity annotation of bounding box, ellipse-wise, pixel-wise. (e),(f),(g) represent the prediction result of bounding box, ellipse-wise, pixel-wise by UNet.

TABLE IV. SEMANTIC SEGMENTATIC EXPERIEMNT RESURT

| Model | Accuracy | Dice | Nuclei recall | Cancer recall |
|---|---|---|---|---|
| **UNet + pw** | **95.79%** | **0.94** | **95.45%** | **75.96%** |
| UNet + el | 92.44% | 0.89 | 93.43% | 74.54% |
| UNet + bd | 89.58% | 0.86 | 94.94% | 75.30% |
| **FPN + pw** | **95.32%** | **0.93** | **94.33%** | **74.20%** |
| FPN + el | 91.58% | 0.88 | 91.03% | 71.21% |
| FPN + bd | 88.84% | 0.85 | 93.09% | 71.33% |
| **LinkNet + pw** | **94.94%** | **0.93** | 92.24% | **69.94%** |
| LinkNet +el | 91.08% | 0.87 | 90.44% | 67.74% |
| LinkNet + bd | 86.61% | 0.80 | **92.40%** | 63.59% |

## B. Semantic segmentation task

For three kinds of granularity annotation: pixel-wise, bounding box, ellipse-wise, we preform semantic segmentation experiments under three models UNet, FPN, LinkNet. Experiment results are shown in Table 4. In UNet model, the accuracy of pixel-wise annotation is 95.79%, which is 3.3% higher than ellipse-wise annotation and 6.2% higher than bounding box annotation. The accuracy of ellipse-wise annotation increased by 2.9% compared with bounding box annotation. In FPN and LinkNet model, pixel-wise annotation achieves similar results. The DICE score is 0.94, 0.93, 0.93 in three models for pixel-wise annotation, which is also the highest. Among the two metrics of the nuclear recall rate and cancer cell recall, expect for the nuclei recall in LinkNet is lightly lower than bounding box and ellipse-wise granularity. The pixel-wise annotation still obtains the best experiment results. In UNet, pixel-wise granularity annotation gets an accuracy of 95.45% and 75.96%. Experiment results show that for a variety of full convolution network, finer-grained annotation can help model extract more accurate features from original pathological slides and improve the performance of the model.

As shown in Fig. 4, for the mask generated by the three granularity, we can clearly see from the prediction results that pixel-wise annotation can clearly depict the outline of all nuclei and the level of the corresponding cancer cells. That is to say, pixel-wise annotation can help deep learning model extract accurate phenotypic information from pathological slides. Accurate acquisition of phenotypic information provides an accurate explanation of the model prediction results, which can help pathologists to understand the prediction result based on which part on slide and improve the interpretability of the model.

## CONCLUSION AND FUTURE WORK

This paper has verified that Finer-grained annotation plays an important role in improving the performance of deep learning model for histopathological image classification and semantic segmentation. In classification, finer-grained annotation can help deep learning model extract more accurate features from pathological slides, improve the recognition rate of different types of cancer. In semantic segmentation, finer-grained annotation can help deep learning model extract more accurate visual phenotype, such as the outline, morphology and color of nucleus according to the surrounding context. The deep learning model trained by finer-grained annotations did not only assist pathologists to perform pathological examination, but also intuitively show the relevant location and grading information of cancer tissue to pathologists, improving the interpretability of deep learning model. Accurate phenotypic information can help to automatically generate pathological diagnosis reports from pathological slide combining with relevant knowledge of natural language processing, promote the development of automatic medical treatment, and contribute to the study of genotype-phenotype association.

## ACKNOWLEDGMENT

This work has been supported by the National Key Research and Development Program of China [2018YFC0910404]; National Natural Science Foundation of China [61772409]; the consulting research project of the Chinese Academy of Engineering (The Online and Offline Mixed Educational Service System for "The Belt and Road" Training in MOOC China); Project of China Knowledge Centre for Engineering Science and Technology; the Innovation Team from the Ministry of Education [IRT\_17R86]; the Innovative Research Group of the National Natural Science Foundation of China [61721002]; and Professor Chen Li' s Recruitment Program for Young Professionals of "The Thousand Talents Plan".